\documentstyle[12pt]{article}
\def\theequation{\arabic{section}.\arabic{equation}}
\begin{document}
\newcommand{\ti}{\theta^{1,0\,\und i}}
\newcommand{\tl}{\theta^{1,0\,\und l}}
\newcommand{\tk}{\theta^{1,0\,\und k}}
\newcommand{\tx}{\theta^{1,0}_{\und i}}
\newcommand{\ta}{\theta^{0,1\,\und a}}
\newcommand{\tb}{\theta^{0,1\,\und b}}
\newcommand{\tc}{\theta^{0,1\,\und c}}
\newcommand{\ty}{\theta^{0,1}_{\und a}}
\newcommand{\da}{D^{2,0}}
\newcommand{\db}{D^{0,2}}
\newcommand{\du}{D^{0,0}_u}
\newcommand{\dv}{D^{0,0}_v}
\newcommand{\qa}{q^{1,0\;\und a}}
\newcommand{\qb}{q^{0,1\;\und i}}
\newcommand{\qc}{q^{\und i\;\und a}}
\newcommand{\nn}{\nonumber}
\newcommand{\be}{\begin{equation}}
\newcommand{\bea}{\begin{eqnarray}}
\newcommand{\eea}{\end{eqnarray}}
\newcommand{\ee}{\end{equation}}
\newcommand{\eps}{\varepsilon}
\newcommand{\und}{\underline}
\newcommand{\p}[1]{(\ref{#1})}

\begin{titlepage}
\begin{flushright}
JINR-E2-2004-150 \\
hep-th/0409236
\end{flushright}
\vskip 1.0truecm
\begin{center}
{\large \bf DIVERSE ${\cal N}{=}(4,4)$ TWISTED MULTIPLETS \\
\vspace{0.3cm}
IN ${\cal N}{=}(2,2)$ SUPERSPACE}
\vglue 1.5cm
{\bf E. Ivanov and A. Sutulin}
\vglue 1.5cm
{\it Bogoliubov Laboratory of Theoretical Physics,
JINR,\\
141 980 Dubna, Moscow Region, Russia}\\
\vspace{0.2cm}
{\tt eivanov,sutulin@thsun1.jinr.ru}
\end{center}

\begin{abstract}
We describe four different types of the ${\cal N}{=}(4,4)$ twisted 
supermultiplets in two-dimensional ${\cal N}{=}(2,2)$ superspace 
${\bf R}^{(1,1|2,2)}$. All these multiplets are presented by a pair 
of chiral and twisted chiral superfields and differ in 
the transformation properties under an extra hidden ${\cal N}{=}(2,2)$ 
supersymmetry. The sigma model ${\cal N}{=}(2,2)$ superfield Lagrangians 
for each type of the ${\cal N}{=}(4,4)$ twisted supermultiplet are
real functions subjected to some differential constraints implied by 
the hidden supersymmetry. We prove that the general sigma model action, 
with all types of ${\cal N}{=}(4,4)$ twisted multiplets originally included, 
is reduced to a sum of sigma model actions for separate types. 
An interaction between the multiplets of different sorts is possible only 
through the appropriate mass terms, and only for those multiplets 
which belong to the same `self-dual' pair.
\end{abstract}
\end{titlepage}

\section{Introduction}

An important class of $2D$ supersymmetric sigma models is constituted by 
${\cal N}{=}(2,2)$ and ${\cal N}{=}(4,4)$ supersymmetric models with 
torsionful bosonic target manifolds and two independent mutually commuting 
left and right complex structures~\cite{IK, GHR, HP, IKL, STP, L}\,.
These models and, in particular, their group manifold WZNW representatives 
can provide non - trivial backgrounds for $4D$ superstrings~\cite{SSTP, RASS, KKL}
and be relevant to $2D$ black holes in stringy context~\cite{RSS, CHS}\,.
Manifestly supersymmetric formulations of ${\cal N}{=}(2,2)$ models in terms 
of chiral and twisted chiral ${\cal N}{=}(2,2)$ superfields, as well as 
in terms of semi -- chiral superfields, have been studied 
in~\cite{GHR, BLR, ST1, ST2, GMST, BSLG}\,.
For ${\cal N}{=}(4,4)$ models with commuting structures there also 
exist manifestly supersymmetric {\it off -- shell} formulations in 
the projective~\cite{RASS, RSS, LIR, IKR} and ordinary~\cite{GI,G1} 
${\cal N}{=}(4,4)$\,, $2D$ superspaces. The basic object of these 
formulations  is the ${\cal N}{=}(4,4)$ twisted multiplet.
In ${\cal N}{=}(2,2)$\,, $2D$ superspace, the latter amounts to a pair 
of chiral and twisted-chiral supefields.

On the other hand, an adequate framework for $4D$ theories with extended 
supersymmetry, e.g., with ${\cal N}{=}2$ one, is provided by the harmonic 
superspace (HSS)  approach~\cite{HSS, book}\,. The ${\cal N}{=}(4,4)$\,, 
$2D$ sigma models which  can be obtained via a direct dimensional reduction 
of ${\cal N}{=}2$\,, $4D$  sigma models constitute a subclass in a more 
general variety of ${\cal N}{=}(4,4)$\,,  $2D$ sigma models.
Indeed, their bosonic target manifolds are hyper -- K\"ahler or 
quaternionic -- K\"ahler, and so are torsionless and exhibit only one set of
 complex structures. In order to describe $2D$ supersymmetric theories 
with torsion one needs a more general type of the harmonic superspace as 
compared to the one  which is of use in the $4D$ case. This new type of 
harmonic superspace,  the $SU(2) \times SU(2)$ bi-harmonic superspace, 
has been constructed in our paper~\cite{IS}\,. Its key feature is 
the presence of two independent sets of harmonic variables which are 
associated with two mutually commuting  automorphism $SU(2)$ groups in 
the left and right light--cone sectors. 
In~\cite{IS} we showed  how to describe one type of the ${\cal N}{=}(4,4)$ 
twisted supermultiplets in the $SU(2) \times SU(2)$ bi-harmonic superspace
and wrote the most general {\it off--shell} action of this multiplet as 
an integral over an analytic subspace of the full superspace. 
This action corresponds to a general ${\cal N}{=}(4,4)$ supersymmetric sigma 
models  with torsion and mutually commuting sets of left and right complex 
structures. \footnote{Some other off--shell ${\cal N}{=}(4,4)$ multiplets 
and the conformal ${\cal N}{=}(4,4)$ supergravity were presented in the 
$SU(2) \times SU(2)$ HSS  in Refs.~\cite{IS2} and ~\cite{bi1}\,, respectively.}
Later on, a more general class of {\it off -- shell} torsionful
${\cal N}{=}(4,4)$ sigma model actions with non-commuting left and right 
complex structures was constructed within the same $SU(2) \times SU(2)$ 
bi-harmonic superspace approach~\cite{EI, EI1}.

In most of previous cases the general sigma model actions with ${\cal N}{=}(4,4)$ 
supersymmetry were written for arbitrary number of twisted multiplets of 
one fixed type, i.e. for those having the same transformation properties
under the $SO(4)_L\times SO(4)_R$ automorphism group of ${\cal N}{=}(4,4)$\,, 
$2D$ super Poincar\'e algebra. In ${\cal N}{=}(2,2)$ superspace, this basically 
means that each pair of chiral and twisted-chiral superfields one deals with  
has the same transformation properties under the extra ${\cal N}{=}(2,2)$ supersymmetry.
On the other hand, as we shall explicitly show in this paper, the extra supersymmetry 
can be realized in {\it different} ways on {\it different} such pairs. 
As found in~\cite{IEA,GK,IS1}\,, in fact there are four essentially distinct types 
of the twisted ${\cal N}{=}(4,4)$ multiplets (up to additional twists related to 
space-time parities~\cite{GK, GRana, GRana1})\,. The basic distinction between them,
in the ${\cal N}{=}(4,4)$ superfield formulations, is the different realization
of the above automorphism group~\cite{GK, IS1}\,. One of the purposes of 
the present paper is to demonstrate that it is a freedom in defining 
the extra ${\cal N}{=}(2,2)$ supersymmetry transformations on a pair of chiral 
and twisted chiral ${\cal N}{=}(2,2)$ superfields, which is responsible for
this diversity of ${\cal N}{=}(4,4)$ twisted multiplets from the 
${\cal N}{=}(2,2)$\,, $2D$ superspace perspective.

In our previous paper~\cite{IS1}\,, the description of four basic different
types of the ${\cal N}{=}(4,4)$ twisted multiplets in the $SU(2) \times SU(2)$ 
bi-harmonic superspace has been presented and general sigma model actions for all 
of them have been constructed. We also discussed the special case of 
the $SU(2) \times SU(2)$ group manifold WZNW sigma model for one of these multiplets. 
We have shown that the general sigma model action of any pair of different 
multiplets is reduced to a sum of sigma model actions of separate multiplets.

In many aspects, the ${\cal N}{=}(2,2)$ superspace formulations of 
${\cal N}{=}(4,4)$ supersymmetric models are more transparent than 
the ${\cal N}{=}(4,4)$ superspace ones. In the present paper we give 
the description of different types of twisted multiplets in 
the ${\cal N}{=}(2,2)$\,, $2D$ superspace ${\bf R}^{(1,1|2,2)}$\,.
This ${\cal N}{=}(2,2)$ superspace approach is used to unravel some proofs
and conclusions of Ref.~\cite{IS1}\,.

In Sect.~2 we review how ${\cal N}{=}(4,4)$ twisted multiplets
are described in the conventional ${\cal N}{=}(4,4)$\,, $2D$ superspace 
${\bf R}^{(1,1|4,4)}$\,. Then, in Sect.~3, we reformulate the irreducibility 
conditions for these multiplets in ${\cal N}{=}(2,2)$\,, $2D$ superspace 
${\bf R}^{(1,1|2,2)}$\, and show, keeping only one ${\cal N}{=}(2,2)$ 
supersymmetry manifest, that in all cases they are equivalent to 
the chirality and twisted chirality conditions for the pair of constituent 
${\cal N}{=}(2,2)$ superfields. The difference between various 
${\cal N}{=}(4,4)$ twisted multiplets proves to be attributed to 
a difference in the transformation properties of these pairs with 
respect to the extra hidden ${\cal N}{=}(2,2)$ supersymmetry.
The actions for the twisted multiplets are constructed as ${\cal N}{=}(2,2)$ 
superspace integrals of some real functions of the constituent 
${\cal N}{=}(2,2)$ superfields. These functions are
subjected to the appropriate differential constraints implied by the hidden
supersymmetry. In Sect.~4 we examine the ${\cal N}{=}(2,2)$ superspace 
actions of two types. The Lagrangians in these actions depend either on 
the ${\cal N}{=}(4,4)$ twisted multiplets belonging
to a `self-dual' pair, or on those from different such pairs.
\footnote{The definition of `dual' twisted multiplets is given in~\cite{G1, GK} 
and~\cite{IS1} (see also the end of Sect.~2 below)\,.} 
We find that in both cases the sigma model-type actions are reduced to a sum of
such actions for separate multiplets, and this property extends to the cases
when a larger number of non-equivalent twisted multiplets is originally involved.
This confirms the analogous result of Ref.~\cite{IS1} obtained within
the bi-harmonic ${\cal N}{=}(4,4)$ superspace approach. We also show that 
the only possibility to gain an interaction between different twisted multiplets 
is to add ${\cal N}{=}(4,4)$ supersymmetric potential (or mass) terms 
mixing up multiplets from the same `self-dual' pair. The multiplets from 
different such pairs cannot interact at all, once again in the full agreement 
with Ref.~\cite{IS1}\,. Our results are summarized in the concluding Sect.~5.

\setcounter{equation}{0}

\section{Twisted Multiplets in ${\cal N}{=}(4,4)$\,, $2D$ Superspace}

We begin by recapitulating some necessary facts about ${\cal N}{=}(4,4)$\,, $2D$
supersymmetry, basically following our papers~\cite{IS, IS1}\,.
In the light-cone parametrization, the standard real ${\cal N}{=}(4,4)$\,, $2D$ 
superspace is defined as the set of the light-cone coordinates
\be
{\bf R}^{(1,1|4,4)} = (Z_L\,, Z_R) =
(\,z^{++}\,, \theta^{+i\, \und k}\,, z^{--}\,, \theta^{-a\, \und b}\,)\,.
\label{4R}
\ee
Here $+$\,,$-$ are light-cone indices and $i$\,, $\und k$\,, $a$\,, $\und b$
are doublet indices of four commuting $SU(2)$\, groups which constitute
the full automorphism group $SO(4)_L\times SO(4)_R$ of ${\cal N}{=}(4,4)$\,, 
$2D$ Poincar\'e superalgebra. The anticommutators of the corresponding spinor
derivatives read
\be
\{\,D_{i\, \und k}\,, D_{j\, \und l} \,\} = 2i\, \eps_{i\,j}\, \eps_{\und k\, \und l}\,
\partial_{++}\,, \quad
\{\,D_{a\, \und b}\,, D_{c\, \und d} \,\} = 2i\, \eps_{a\,c}\, \eps_{\und b\, \und d}\,
\partial_{--}
\label{alg}
\ee
where
\bea
D_{i\, \und k} = \frac{\partial}{\partial \theta^{i\, \und k}} +
i\theta_{i\, \und k}\,\partial_{++}\,, \qquad
D_{a\, \und b} = \frac{\partial}{\partial \theta^{a\, \und b}} +
i\theta_{a\, \und b}\,\partial_{--}\,.
\eea
We shall use both the quartet notation for spinor derivatives and
Grassmann coordinates and their complex doublet form.
To make the presentation more concise, we omit the light-cone indices
in the quartet notation. The precise relation between these two
notations is as follows
\bea
&&
(\theta^{+i}\,, \bar \theta^{+i}) \equiv \theta^{i\, \und k}\,, \quad
(\theta^{i\, \und k})^{\dagger} = \eps_{i\,l}\, \eps_{\und k\, \und n}\,
\theta^{l\, \und n}\,, \nn\\
&&
(\theta^{-a}\,, \bar \theta^{-a}) \equiv \theta^{a\, \und b}\,, \quad
(\theta^{a\, \und b})^{\dagger} = \eps_{a\,c}\, \eps_{\und b\, \und d}\,
\theta^{c\, \und d}\,, \nn\\
&&
(D_{+i}\,, \bar D_{+i}) \equiv D_{i\, \und k}\,, \quad
(D_{i\, \und k})^{\dagger} = - \eps^{i\,l}\, \eps^{\und k\, \und n}\, D_{l\, \und n}\,, \nn\\
&&
(D_{-a}\,, \bar D_{-a}) \equiv D_{a\, \und b}\,, \quad
(D_{a\, \und b})^{\dagger} = - \eps^{a\,c}\, \eps^{\und b\, \und d}\, D_{c\, \und d}\,.
\label{quart}
\eea
Here the symbol $\dagger$ means the complex conjugation.

The torsionful ${\cal N}{=}(4,4)$ supersymmetric sigma models
can be formulated either in terms of pairs of ${\cal N}{=}(2,2)$ chiral 
and twisted chiral superfields,
or via the properly constrained ${\cal N}{=}(4,4)$ superfields. 
Both superfield sets
represent off-shell twisted ${\cal N}{=}(4,4)$ multiplets.

In ${\cal N}{=}(2,2)$\,, $2D$ superspace these models are described
by an action which is an integral of some real potential~\cite{GHR,RASS,RSS} 
depending on several pairs of ${\cal N}{=}(2,2)$
chiral and twisted chiral superfields. This function satisfies
a set of differential constraints which ensure the off-shell invariance of 
the action under an extra ${\cal N}{=}(2,2)$ supersymmetry. The latter mixes up 
chiral superfields with the twisted chiral ones and, together with 
the manifest ${\cal N}{=}(2,2)$ supersymmetry,
constitutes the full ${\cal N}{=}(4,4)$ supersymmetry.

On the other hand, in the superspace ${\bf R}^{(1,1|4,4)}$ the same 
${\cal N}{=}(4,4)$ twisted multiplet is described by a real quartet superfield 
subjected to the proper irreducibility
constraints~\cite{IK,GHR,IKL,Sig}\,. These constraints reduce the
full set of the field components of this superfield to the {\it off--shell} 
field content $(\bf 8 + 8 )$ of the twisted multiplet.

In most previous studies, the general sigma model actions with 
${\cal N}{=}(4,4)$ supersymmetry, both in the ${\bf R}^{(1,1|2,2)}$~\cite{GHR} 
and ${\bf R}^{(1,1|4,4)}$~\cite{IS} superspace approaches, were 
constructed in terms of only {\it one} kind of ${\cal N}{=}(4,4)$ twisted multiplet.
However, in fact there are
few types of these multiplets which differ in the transformation properties
of their component fields with respect to the full R-symmetry (or automorphism) 
group $SO(4)_L\times SO(4)_R$ of ${\cal N}{=}(4,4)$\,, $2D$ Poincar\'e superalgebra.
This degeneracy of the twisted multiplets was first noticed in~\cite{GI, IEA, GK}\,.

Following Refs.~\cite{IEA,IS1}\,, one can consider four types of twisted multiplets
in the superspace ${\bf R}^{(1,1|4,4)}$\,, in accord with the four
possibilities to pair the doublet indices of various $SU(2)$ factors
of the left and right $SO(4)$ subgroups in the full $SO(4)_L\times SO(4)_R$
automorphism group
\be
\hat q^{\,i\,a}\,,\quad \hat q^{\,i\, \und a}\,,\quad
\hat q^{\,\und i\, a}\,,\quad \hat q^{\,\und i\, \und a}\,.
\label{tw}
\ee
The reality properties of these multiplets in ${\bf R}^{(1,1|4,4)}$\, are
defined by the rules
\bea
&&
(\hat q^{\,i\,a})^{\dagger} =
\eps_{i\,k}\, \eps_{a\,b}\, \hat q^{\,k\,b}\,, \qquad
(\hat q^{\,i\, \und a})^{\dagger} =
\eps_{i\,k}\, \eps_{\und a\, \und b}\, \hat q^{\,k\, \und b}\,, \nn\\
&&
(\hat q^{\,\und i\, a})^{\dagger} =
\eps_{\und i\, \und k}\, \eps_{a\,b}\, \hat q^{\,\und k\, b}\,, \qquad
(\hat q^{\,\und i\, \und a})^{\dagger} =
\eps_{\und i\, \und k}\, \eps_{\und a\, \und b}\, \hat q^{\,\und k\, \und b}\,.
\label{real}
\eea
The irreducibility constraints leaving just total of $(\bf 8+8)$\, independent
components in every such superfield read as
\be
D^{(\,k\, \und k}\, \hat q^{\,i\,)\,a} = 0\,, \qquad
D^{(\,b\, \und b}\, \hat q^{\,i\,a\,)} = 0\,,
\label{con1}
\ee
\be
D^{(\,k\, \und k}\, \hat q^{\,i\,)\, \und a} = 0\,, \qquad
D^{b\, (\,\und b}\, \hat q^{\,i\, \und a\,)} = 0\,,
\label{con2}
\ee
\be
D^{k\, (\,\und k}\, \hat q^{\,\und i\,)\, a} = 0\,, \qquad
D^{(\,b\, \und b}\, \hat q^{\,\und i\,  a\,)} = 0\,,
\label{con3}
\ee
\be
D^{k\, (\,\und k}\, \hat q^{\,\und i\,)\, \und a} = 0\,, \qquad
D^{b\, (\,\und b}\, \hat q^{\,\und i\, \und a\,)} = 0\,
\label{con4}
\ee
where $(\,)$ means the symmetrization in the appropriate indices.
The natural description of these twisted multiplets is achieved~\cite{IS,IS1} 
in ${\cal N}{=}(4,4)$\,, $SU(2)\times SU(2)$ HSS with the double sets
of harmonic variables. The corresponding general {\it off-shell} sigma model 
actions for each kind of these multiplets can be written in the {\it analytic} 
bi-harmonic superspace~\cite{IS,IS1}\,, which is a subspace of the HSS just mentioned.

As shown in~\cite{IS1}\,, in this analytic superspace one can
generalize the sigma model actions of twisted multiplets of one given type
to the cases when the superfield Lagrangian bears a dependence on two or even 
more different species of such a multiplet. 
The basic goal of the present paper is to rephrase these results
in terms of ${\cal N}{=}(2,2)$ chiral and twisted chiral superfields, making them
more tractable and transparent.

To close this Section, let us recall the definition of the `self-dual' and 
`non-self-dual' pairs of the multiplets from the set \p{tw}\,. The `self-dual' pairs 
are comprised by those superfields which have no $SU(2)$ doublet indices in common, 
i.e., by $(\hat q^{\,i\,a}\,, \hat q^{\,\und i\, \und a})$ and
$(\hat q^{\,i\,\und a}\,, \hat q^{\,\und i\, a})\,$. 
Their distinguishing feature is that the physical bosonic fields of one superfield 
within the given pair have the same $SU(2)$ content as the auxiliary
fields of the other. Any other pair is by definition `non-self-dual'.

\setcounter{equation}{0}

\section{Various Twisted Multiplets in ${\cal N}{=}(2,2)$\,, 2D \break Superspace}

In this Section we show how different types of ${\cal N}{=}(4,4)$ 
twisted multiplets can be described in the ${\cal N}{=}(2,2)$\,, $2D$ 
superspace and how the difference between them manifests itself in 
this ${\cal N}{=}(2,2)$ superspace setting.
We find that each type is represented by a pair of chiral and twisted
chiral ${\cal N}{=}(2,2)$ superfields having, however, different properties 
under the extra ${\cal N}{=}(2,2)$ supersymmetry transformations.
Then we demonstrate that the general ${\cal N}{=}(4,4)$ supersymmetric 
sigma model actions of separate multiplets can be constructed in 
${\cal N}{=}(2,2)$ superspace as integrals of real functions subjected to 
some differential constraints.

\subsection{Supersymmetry transformations and constraints}

Let us pass to the equivalent notation for the ${\bf R}^{(1,1|4,4)}$
Grassmann coordinates and spinor derivatives (\ref{4R})\,,\,(\ref{quart})
\bea
&&
\theta^{+i} = (\theta^+\,, \eta^+)\,, \quad
\bar \theta^+_i = (\bar \theta^+\,, \bar \eta^+)\,, \quad
\theta^{-a} = (\theta^-\,, \xi^-)\,, \quad
\bar \theta^-_a = (\bar \theta^-\,, \bar \xi^-)\\
\label{set}
&&
D_{+i} = (D_+\,, d_+)\,, \quad
\bar D_+^i = (\bar D_+\,, \bar d_+)\,, \quad
D_{-a} = (D_-\,, d_-)\,,\quad
\bar D_-^a = (\bar D_-\,, \bar d_-)\,.
\label{Der}
\eea
In what follows, the coordinates $\theta$'s and spinor derivatives $D$ refer
to the manifest ${\cal N}{=}(2,2)$ supersymmetry,
while the coordinates $\eta$'s, $\xi$'s
and the derivatives $d$ to the extra hidden ${\cal N}{=}(2,2)$ one
(see the Appendix A for the precise relation between the ${\cal N}{=}(4,4)$
superspace covariant derivatives (\ref{quart}) and these
${\cal N}{=}(2,2)$ ones)\,.

The standard real ${\cal N}{=}(2,2)$\,, $2D$ superspace ${\bf R}^{(1,1|2,2)}$
is parametrized by the following set of coordinates
\be
{\bf R}^{(1,1|2,2)} = (z^{++}\,, \theta^+\,, \bar \theta^+\,, 
z^{--}\,, \theta^-\,, \bar \theta^-)\,.
\label{space}
\ee

Now we wish to see what the irreducibility conditions
(\ref{con1})--(\ref{con4}) look like in the superspace  ${\bf R}^{(1,1|2,2)}$\,.
Keeping in mind the reality conditions (\ref{real}) for twisted multiplets
in ${\bf R}^{(1,1|4,4)}\,$, we introduce the following complex
${\cal N}{=}(4,4)$ superfields
\bea
&&
\hat q^{\,1\,1} = {\bf A}\,,\quad
\hat q^{\,1\,2} = {\bf B}\,,\quad
\hat q^{\,2\,1} = - \bar {\bf B}\,,\quad
\hat q^{\,2\,2} = \bar {\bf A}\,, \\
\label{field1}
&&
\hat q^{\,1\, \und 1} = {\bf a},\quad
\hat q^{\,1\, \und 2} = {\bf b}\,, \quad
\hat q^{\,2\, \und 1} = - \bar {\bf b}\,,\quad
\hat q^{\,2\, \und 2} = \bar {\bf a}\,,
\label{field2}
\eea
\be
\hat q^{\,\und 1\, 1} = {\cal A}\,,\quad
\hat q^{\,\und 1\, 2} = {\cal B}\,, \quad
\hat q^{\,\und 2\, 1} = - \bar {\cal B}\,,\quad
\hat q^{\,\und 2\, 2} = \bar {\cal A}\,,
\label{field3}
\ee
\be
\hat q^{\,\und 1\, \und 1} = {\sf A }\,,\quad
\hat q^{\,\und 1\, \und 2} = {\sf B}\,, \quad
\hat q^{\,\und 2\, \und 1} = - \bar {\sf B}\,,\quad
\hat q^{\,\und 2\, \und 2} = \bar {\sf A}\,.
\label{field4}
\ee
Expanding these ${\cal N}{=}(4,4)$ superfields with respect to the extra
Grassmann coordinates $\eta^\pm$\,, $\xi^\pm$ (and their conjugates)
and using the constraints (\ref{con1})--(\ref{con4})\,, one finds that 
only the first components of each ${\cal N}{=}(4,4)$ superfield are 
independent ${\cal N}{=}(2,2)$ superfields defined on the superspace 
${\bf R}^{(1,1|2,2)}$\,.
The ${\cal N}{=}(2,2)$ coefficients of the higher $\eta$\,, $\xi$\, 
monomials are expressed as spinor $D$ derivatives of the coefficients 
associated with the lower-order monomilas.
The leading lowest-order components satisfy the chirality and 
the twisted chirality conditions in ${\bf R}^{(1,1|2,2)}$\,.

E.g. for the multiplet $\hat q^{\,i\,a}$ these conditions read
(see the Appendix A for the analogous conditions for other types 
of twisted multiplets)
\bea
&&
\bar D_+ A = 0\,,\qquad
\bar D_- A = 0\,, \qquad
\bar D_+ B = 0\,,\qquad
D_- B = 0\,, \nn\\
&&
D_+ \bar A = 0\,,\qquad
D_- \bar A = 0\,,\qquad
D_+ \bar B = 0\,,\qquad
\bar D_- \bar B = 0 \,.
\label{chir1}
\eea
Here
\be
A = \left. {\bf A} \right |_{\eta=\xi=0}\,, \quad
 \bar A = \left. \bar {\bf A} \right |_{\eta=\xi=0}\,, \quad
B = \left. {\bf B} \right |_{\eta=\xi=0}\,, \quad
\bar B = \left. \bar {\bf B} \right |_{\eta=\xi=0}\,.
\label{f}
\ee

To see how the difference between non-equivalent ${\cal N}{=}(4,4)$ twisted 
multiplets manifests itself in this approach, let us explicitly quote 
the extra Grassmann coordinate expansions for those
${\cal N}{=}(4,4)$ superfields which have as the first component
the ${\cal N}{=}(2,2)$ chiral field
\bea
&&
{\mbox(i)}\;\;\;\; {\bf A} = A - \bar \eta^+ \bar D_+ \bar B 
+ \bar \xi^- \bar D_- B + ...,\qquad
{\sf A} = {\sf a} + \eta^+ \bar D_+ \bar {\sf b} - \xi^- \bar D_- {\sf b} + ...,\nn\\
&&
{\mbox(ii)}\;\;\;\; {\bf a} = a - \bar \eta^+ \bar D_+ \bar b - \xi^- \bar D_- b + ...,\qquad
{\cal A} = \alpha + \eta^+ \bar D_+ \bar \beta + \bar \xi^- \bar D_- \beta + ...,\nn\\
&&
A = \left. {\bf A} \right |_{\eta=\xi=0}\,, \quad
 a = \left. {\bf a} \right |_{\eta=\xi=0}\,, \quad
\alpha = \left. {\cal A} \right |_{\eta=\xi=0}\,, \quad
{\sf a} = \left. {\sf A} \right |_{\eta=\xi=0}\,.
\label{f1}
\eea

In the superspace ${\bf R}^{(1,1|4,4)}$ the supersymmetry
transformation of the ${\cal N}{=}(4,4)$ superfield $\Phi$ are generated by
differential operators $Q$ the explicit form of which is given in (\ref{gen})
\be
\delta \Phi = i\, \Big (\eps^{\,+k} Q_{+k} - \bar \eps^{\,+}_k \bar Q^k_+
+ \eps^{\,-a} Q_{-a} - \bar \eps^{\,-}_a \bar Q^a_- \Big ) \,\Phi\,.
\label{susy4}
\ee
When we reduce the superspace ${\bf R}^{(1,1|4,4)}$ to its subspace 
${\bf R}^{(1,1|2,2)}$\,, half of supersymmetries become non-manifest 
and their transformations explicitly involve spinor derivatives. 
These extra supersymmetries mix different ${\cal N}{=}(2,2)$ superfields
which are the first components of the ${\cal N}{=}(4,4)$ superfields 
defined in Eqs.~(\ref{f})\,,\,(\ref{f2}) -- (\ref{f4})\,.

Substituting the expansions, such as (\ref{f1})\,,
for each ${\cal N}{=}(4,4)$ superfield
into the transformation law (\ref{susy4}) and singling out the subset
with $k = a =2$ there (the corresponding generators and infinitesimal parameters
are associated just with the hidden ${\cal N}{=}(2,2)$ supersymmetry),
one finds the extra ${\cal N}{=}(2,2)$ supersymmetry transformation laws 
of the first components of these superfields
\bea
&&\delta A = \bar \eps^{\,+} \bar D_+ \bar B - \bar \eps^{\,-} \bar D_- B\,, \qquad \quad
\delta {\sf a} = - \eps^{\,+} \bar D_+ \bar {\sf b} + \eps^{\,-} \bar D_- {\sf b}\,,
\nn\\
&&
\delta \bar A = - \eps^{\,+} D_+ B + \eps^{\,-} D_- \bar B\,, \;\qquad
\delta \bar {\sf a} = \bar \eps^{\,+} D_+ {\sf b} - \bar \eps^{\,-} D_- \bar {\sf b}\,,
\nn\\
&&\delta B = - \bar \eps^{\,+} \bar D_+ \bar A - \eps^{\,-} D_- A\,,\;\qquad
\delta {\sf b} = \eps^{\,+} \bar D_+ \bar {\sf a} + \bar \eps^{\,-} D_- {\sf a}\,,
\nn\\
&&\delta \bar B = \eps^{\,+} D_+ A + \bar \eps^{\,-} \bar D_- \bar A\,, \quad\, \qquad
\delta \bar {\sf b} = - \bar \eps^{\,+} D_+ {\sf a} - \eps^{\,-} \bar D_- \bar {\sf a}\,,
\label{s1}
\eea
\bea
&&
\delta a = \bar \eps^{\,+} \bar D_+ \bar b + \eps^{\,-} \bar D_- b\,, \qquad \qquad
\delta \alpha = - \eps^{\,+} \bar D_+ \bar \beta - \bar \eps^{\,-} \bar D_- \beta\,,
\nn\\
&&\delta \bar a = - \eps^{\,+} D_+ b - \bar \eps^{\,-} D_- \bar b\,, \quad \,\qquad
\delta \bar \alpha = \bar \eps^{\,+} D_+ \beta + \eps^{\,-} D_- \bar \beta\,,
\nn\\
&&\delta b = - \bar \eps^{\,+} \bar D_+ \bar a + \bar \eps^{\,-} D_- a\,, \quad \,\qquad
\delta \beta = \eps^{\,+} \bar D_+ \bar \alpha - \eps^{\,-} D_- \alpha\,,
\nn\\
&&\delta \bar b = \eps^{\,+} D_+ a - \eps^{\,-} \bar D_- \bar a\,, \qquad \qquad
\delta \bar \beta = - \bar \eps^{\,+} D_+ \alpha + \bar \eps^{\,-} \bar D_- \bar \alpha\,.
\label{s2}
\eea
In (\ref{s1}) and (\ref{s2}) we collected, respectively, the transformation laws 
of ${\cal N}{=}(2,2)$ superfields belonging to the `self-dual' pairs of 
the ${\cal N}{=}(4,4)$ twisted multiplets, i.e., $(\hat q ^{\,i\,a}\,, 
\hat q^{\,\und i\, \und a})$ and 
$(\hat q^{\,i\,\und a}\,, \hat q^{\,\und i\, a})\,$.

Looking at the sets of transformation laws (\ref{s1}) and (\ref{s2})\,, 
we come to the conclusion that the only difference in the description 
of various ${\cal N}{=}(4,4)$ twisted multiplets in the superspace 
${\bf R}^{(1,1|2,2)}$ lies in the transformation laws of their chiral 
and twisted chiral ${\cal N}{=}(2,2)$ constituents under the extra 
${\cal N}{=}(2,2)$ supersymmetry. These laws are specific for each multiplet.

Thus, in the superspace ${\bf R}^{(1,1|2,2)}\,$, each type of
the ${\cal N}{=}(4,4)$ twisted multiplets is represented by a pair
of chiral and twisted chiral ${\cal N}{=}(2,2)$ superfields.
Despite this common feature, the extra ${\cal N}{=}(2,2)$ supersymmetry 
is realized differently on the ${\cal N}{=}(2,2)$ superfields 
from non-equivalent twisted multiplets.
For ${\cal N}{=}(2,2)$ superfields of one kind (chiral or twisted chiral)
belonging to different ${\cal N}{=}(4,4)$ twisted multiplets one cannot
{\it simultaneously} bring the hidden ${\cal N}{=}(2,2)$ supersymmetry
transformation laws into the same form by any redefinition of 
the transformation parameters and/or involved superfields.

Note that the full automorphism group $SO(4)_L\times SO(4)_R$ of 
the ${\cal N}{=}(4,4)$\,, $2D$ super Poincar\'e algebra has different 
realizations on diverse ${\cal N}{=}(4,4)$ twisted
multiplets in the ${\cal N}{=}(2,2)$\,, $2D$  superspace formulation, 
precisely as in the ${\cal N}{=}4,4)$ superspace one. 
However, only an $U(1)_L\times U(1)_R$ subgroup of this automorphism group 
is manifest in the ${\cal N}{=}(2,2)$ setting. So it is the difference
in the realizations of hidden ${\cal N}{=}(2,2)$ supersymmetry which 
is the basic distinguishing feature of non-equivalent twisted multiplets 
in the ${\cal N}{=}(2,2)$ superfield description.

\subsection{Action for $\hat q^{\,i\,a}$ multiplet}

The general action of $k$ chiral superfields $C^k$ and $n$
twisted chiral superfields $T^n$ can be written in ${\bf R}^{(1,1|2,2)}$
as an integral of some real function $K$
\be
S_{(2,2)} = \int \mu \, K(C^k\,,\, \bar C^k\,,\, T^n\,,\, \bar T^n)
\label{22}
\ee
where
\be
\mu = d^2x\, d^2 \theta^+ \, d^2 \theta^-
= d^2x\, d\theta^+ \, d\bar \theta^+ \, d\theta^- \, d\bar \theta^-
\ee
is the integration measure. The superpotential $K$ is defined modulo
a generalized K\"ahler transformation
\be
\delta K = f\,(C^k\,,\, T^n) + g\,(C^k\,,\, \bar T^n)
+ \bar f\,(\bar C^k\,, \, \bar T^n) + \bar g\,(\bar C^k\,, \, T^n)\,.
\ee
When this action describes a theory with only ${\cal N}{=}(2,2)$ supersymmetry, 
the numbers of chiral and twisted chiral superfields are not obliged 
to coincide, $k \neq n\,$. One can also add, without introducing any 
central charges, scalar potential interaction terms to Eq.~(\ref{22})\,,
which involve two holomorphic functions $P_1(C^k)$\,, $P_2(T^n)$
\be
S^{\,pot}_{(2,2)} \,=\, i\, m \int \mu\,
\Big \{\,(\bar \theta^+ \bar \theta^-) P_1\,(C^k)
+ (\theta^+ \theta^-) \bar P_1\, (\bar C^k)
+ (\bar \theta^+ \theta ^-) P_2\,(T^n)
+ (\theta^+ \bar \theta^-) \bar P_2\,(\bar T^n)
\Big \}\,.
\label{pot}
\ee
Despite the presence of explicit $\theta$'s, these terms are invariant 
under the manifest ${\cal N}{=}(2,2)$ supertranslations as a consequence of 
the chirality and twisted-chirality constraints for the corresponding 
superfields. After elimination of the auxiliary fields in the sum of 
the actions \p{22} and \p{pot}\,, the resulting component actions acquire
some scalar potentials of physical bosonic fields.

Let us firstly assume that the chiral $C$
(antichiral $\bar C$\,) and twisted chiral $T$ (twisted antichiral $\bar T$\,)
superfields comprise one kind of ${\cal N}{=}(4,4)$ twisted multiplet.
Then, requiring the action (\ref{22}) to possess
an extra ${\cal N}{=}(2,2)$ supersymmetry implies, first, that the numbers 
of chiral and twisted chiral superfields must coincide, $k=n$, and, second,
gives rise to some differential constraints on the function $K \,$.
These constraints read~\cite{GHR}
\be
\frac{\partial^2 K}{\partial C^l \partial \bar C^k}
- \frac{\partial^2 K}{\partial C^k \partial \bar C^l} = 0\,,\qquad
\frac{\partial^2 K}{\partial T^l \partial \bar T^k}
- \frac{\partial^2 K}{\partial T^k \partial \bar T^l} = 0\,,
\label{AA}
\ee
\be
\frac{\partial^2 K}{\partial C^{(\,k} \partial \bar C^{l\,)}}
+ \frac{\partial^2 K}{\partial T^{(\,l} \partial \bar T^{k\,)}} = 0\,.
\label{AAA}
\ee
In particular, if the potential $K$ involves only one chiral and 
one twisted chiral multiplet $(m=1)\,$, the set of constraints 
(\ref{AA})\,, (\ref{AAA}) amounts to the single
four-dimensional Laplace equation
\be
\frac{\partial^2 K}{\partial C \partial \bar C}
+ \frac{\partial^2 K}{\partial T \partial \bar T} = 0\,.
\label{AA1}
\ee

Thus for any real $K$ obeying (\ref{AA})\,, (\ref{AAA}) the action
\be
S_{(4,4)} = \int \mu \, K(C^m\,,\, \bar C^m\,,\, T^m\,,\, \bar T^m)
\ee
is ${\cal N} = (4,4)$ supersymmetric.

It can be also checked that, for the potential (or mass) terms (\ref{pot})
to preserve ${\cal N}{=}(4,4)$ supersymmetry, the functions $P_1$\,,\,$P_2$ 
should be linear in the corresponding superfields
\be
S^{\,m}_{(4,4)} \,=\, i\, m \int \mu\, \Big \{\,
(\bar \theta^+ \bar \theta^-)\, C
+ (\theta^+ \theta^-)\, \bar C
+ (\bar \theta^+ \theta ^-)\, T
+ (\theta^+ \bar \theta^-)\, \bar T
\Big \}\,.
\label{mass1}
\ee
Although in the component form the action \p{mass1} contains only terms 
linear in the auxiliary fields, the elimination of the latter in the full 
action including also a sigma-model part can give rise to non-trivial 
scalar potentials of physical bosonic fields (provided that the target 
space bosonic metric is non-trivial)~\cite{IK,IKL,GI,IS}\,.

This consideration can be extended to any other type of
${\cal N}{=}(4,4)$ twisted multiplet. If the action bears dependence only on
those chiral and twisted chiral ${\cal N}{=}(2,2)$
superfields which comprise the same type of ${\cal N}{=}(4,4)$ twisted multiplets,
the extra ${\cal N}{=}(2,2)$ supersymmetry differential constraints for 
the relevant potential $K$ have the same form (\ref{AA})\,, (\ref{AAA})\,, 
whatever the twisted multiplet is. The structure of the off-shell potential terms
in this case is also uniquely fixed by the requirement of extra ${\cal N}{=}(2,2)$ 
supersymmetry. They are given by a sum of the actions (\ref{mass1})\,.

\setcounter{equation}{0}

\section{Actions for a Pair of Twisted Multiplets}

\subsection{Preface: the free actions}

In the previous Section we have found that the chiral and twisted chiral
superfields forming one or another type of the ${\cal N}{=}(4,4)$
twisted multiplet have different transformation properties under the hidden
${\cal N}{=}(2,2)$ supersymmetry.
Now we are going to construct, in the superspace ${\bf R}^{(1,1|2,2)}\,$,
the supersymmetric sigma model actions of two types,
with the dependence on either a `non-self-dual' or `self-dual' pair
of the ${\cal N}{=}(4,4)$ twisted multiplets. We will show that
in both cases the corresponding sigma model actions are
reduced to a sum of sigma model actions for the separate twisted multiplets.
The results presented below are in a full agreement with those obtained
in Ref.~\cite{IS1} within the harmonic superspace approach.

Before turning to the general case, let us recall the HSS description of 
two instructive examples of the action with two twisted multiplets,
viz. the actions which are bilinear in the corresponding superfields~\cite{IS1}\,.
The first option is the general quadratic actions
depending only on one kind of the twisted multiplet and \'a priori including
some harmonic constants. Requiring it to be ${\cal N}{=}(4,4)$ supersymmetric
leads to the conditions on these constants and, as a result, the corresponding
actions are reduced to the relevant free actions.
In the second case, the bilinear actions involve different sorts of 
the twisted multiplets. The inspection of such actions in HSS~\cite{IS1} leads to
the conclusion that the requirement of invariance under the ${\cal N}{=}(4,4)$
supersymmetry implies them to vanish.

These results can be easily reproduced in the ${\bf R}^{(1,1|2,2)}$ superspace
formalism in terms of chiral and twisted chiral superfields. It is easy to show
that the bilinear sigma model action which contains chiral and twisted chiral 
superfields comprising one sort of the twisted ${\cal N}{=}(4,4)$ multiplet,
\be
S_{(4,4)}^{free} = \int \mu\, (C^m \bar C^m - T^m \bar T^m)\,,
\label{free}
\ee
is equivalent to the ${\cal N}{=}(4,4)$ supersymmetric free actions
of such twisted multiplets, while the ${\bf R}^{(1,1|2,2)}$ actions
constructed bilinear in chiral and twisted chiral superfields from different 
types of the ${\cal N}{=}(4,4)$ twisted multiplets are vanishing as 
a consequence of extra ${\cal N}{=}(2,2)$ supersymmetry. 
Note, that the relative sign between two terms in (\ref{free}) is 
uniquely fixed by hidden ${\cal N}{=}(2,2)$ supersymmetry (the component 
Lagrangian is positive-definite despite this sign minus in \p{free})\,.

Below we shall repeat in ${\bf R}^{(1,1|2,2)}$ our general analysis
of sigma model actions both for the `non-self-dual' and `self-dual' pairs
of twisted multiplets~\cite{IS1}\,. The unique form of the free action \p{free} 
and the property that the actions bilinear in the ${\cal N}{=}(2,2)$ superfields 
from different kinds of ${\cal N}{=}(4,4)$ twisted multiplets are vanishing 
will follow from this general analysis.

\subsection{Action for non-dual twisted multiplets}

We start from the action for the multiplets $\hat q^{\,i\,a}$ and 
$\hat q^{\,i\, \und a}$ belonging to different `self-dual' pairs. 
It is given as an integral of some real function $K$ over 
the superspace ${\bf R}^{(1,1|2,2)}$
\be
\label{Lagr1}
S_{(4,4)} = \int \mu \, K(A\,,\, \bar A\,,\, B\,,\, 
\bar B\,,\, a\,,\, \bar a\,,\, b\,,\, \bar b)\,.
\ee
Since $K$ is a function of ${\cal N}{=}(2,2)$ superfields, the action
(\ref{Lagr1}) is evidently invariant under the manifest ${\cal N}{=}(2,2)$ 
supersymmetry. It is also invariant under generalized K\"ahler gauge 
transformations
\be
\delta K = f\,(A\,, B\,, a\,, b) + g\,(A\,, \bar B\,, a\,, \bar b)
+ \bar f\,(\bar A\,, \bar B\,, \bar a\,, \bar b) 
+ \bar g\,(\bar A\,, B\,, \bar a\,, b)\,.
\label{Gauge1}
\ee
Using the chirality and twisted chirality conditions for the involved 
superfields and the definition of the integration measure on 
the superspace ${\bf R}^{(1,1|2,2)}$\,,
it is  easy to show that the gauge functions in \p{Gauge1} indeed
do not contribute into the ${\cal N}{=}(2,2)$ superfield action.

We require this action to admit an extra ${\cal N}{=}(2,2)$
supersymmetry which is realized on superfields according to 
Eqs. (\ref{s1})\,, (\ref{s2})\,. This requirement amounts to some 
additional constraints on $K\,$. The general condition of 
the invariance of the action \p{Lagr1} under these transformations is
as follows
\bea
\delta S_{(4,4)} \!&=&\! \int \mu\, \Big \{\,
\eps^{\,+} D_+ F - \bar \eps^{\,+} \bar D_+ \bar F +
\eps^{\,+} \bar D_+ G - \bar \eps^{\,+} D_+ \bar G \nn\\
\!&+&\! \eps^{\,-} D_- H - \bar \eps^{\,-} \bar D_- \bar H +
\eps^{\,-} \bar D_- P - \bar \eps^{\,-} D_- \bar P\, \Big \}\,
\label{deltaS}
\eea
where, for the moment, the functions in the r.h.s. are arbitrary.
Explicitly computing the variation $\delta S_{(4,4)}$ and comparing 
both parts of \p{deltaS}\,, as the result of equating the coefficients 
before independent infinitesimal parameters we obtain
\bea
\eps^{\,+} \quad \Rightarrow \quad
\frac{\partial K}{\partial \bar B} = \frac{\partial F}{\partial A}\,, \quad
\frac{\partial K}{\partial \bar A} = - \frac{\partial F}{\partial B}\,, \quad
\frac{\partial K}{\partial \bar b} = \frac{\partial F}{\partial a}\,, \quad
\frac{\partial K}{\partial \bar a} = - \frac{\partial F}{\partial b}\,,
\nn\\
\bar \eps^{\,+} \quad \Rightarrow \quad
\frac{\partial K}{\partial B} = \frac{\partial \bar F}{\partial \bar A}\,, \quad
\frac{\partial K}{\partial A} = - \frac{\partial \bar F}{\partial \bar B}\,, \quad
\frac{\partial K}{\partial b} = \frac{\partial \bar F}{\partial \bar a}\,, \quad
\frac{\partial K}{\partial a} = - \frac{\partial \bar F}{\partial \bar b}\,,
\label{x1}
\eea
\bea
\eps^{\,-}  \quad \Rightarrow \quad
\frac{\partial K}{\partial B} = - \frac{\partial H}{\partial A}\,, \quad
\frac{\partial K}{\partial \bar A} = \frac{\partial H}{\partial \bar B}\,, \quad
\frac{\partial K}{\partial \bar b} = - \frac{\partial P}{\partial \bar a}\,, \quad
\frac{\partial K}{\partial a} = \frac{\partial P}{\partial b}\,,
\nn\\
\bar \eps^{\,-}  \quad \Rightarrow \quad
\frac{\partial K}{\partial \bar B} = - \frac{\partial \bar H}{\partial \bar A}\,, \quad
\frac{\partial K}{\partial A} = \frac{\partial \bar H}{\partial B}\,, \quad
\frac{\partial K}{\partial b} = - \frac{\partial \bar P}{\partial a}\,, \quad
\frac{\partial K}{\partial \bar a} = \frac{\partial \bar P}{\partial \bar b}\,.
\label{x2}
\eea
Along with these constraints, we find that the functions in the r.h.s. 
of (\ref{deltaS}) should obey the additional analyticity-type conditions
\be
\eps^{\,+} \quad \Rightarrow \quad G = 0\,, \qquad
\bar \eps^{\,+} \quad \Rightarrow \quad \bar G = 0\,,
\ee
\bea
\eps^{\,-} \quad \Rightarrow \quad
\frac{\partial H}{\partial a} = 0\,, \qquad
\frac{\partial H}{\partial \bar b} = 0\,, \qquad
\frac{\partial P}{\partial \bar A} = 0\,, \qquad
\frac{\partial P}{\partial B} = 0\,,
\nn\\
\bar \eps^{\,-} \quad \Rightarrow \quad
\frac{\partial \bar H}{\partial \bar a} = 0\,, \qquad
\frac{\partial \bar H}{\partial b} = 0\,, \qquad
\frac{\partial \bar P}{\partial A} = 0\,, \qquad
\frac{\partial \bar P}{\partial \bar B} = 0\,.
\label{x3}
\eea
The integrability conditions for $K$ following from 
the constraints (\ref{x1}) read
\bea
&& \frac{\partial^2 K}{\partial A \partial \bar A}
+ \frac{\partial^2 K}{\partial B \partial \bar B} = 0\,, \qquad
\frac{\partial^2 K}{\partial a \partial \bar a}
+ \frac{\partial^2 K}{\partial b \partial \bar b} = 0\,, \nn\\
&&  \frac{\partial^2 K}{\partial a \partial \bar B}
- \frac{\partial^2 K}{\partial \bar b \partial A} = 0\,, \qquad
\frac{\partial^2 K}{\partial \bar a \partial B}
- \frac{\partial^2 K}{\partial b \partial \bar A} = 0\,, \nn \\
&&  \frac{\partial^2 K}{\partial a \partial \bar A}
+ \frac{\partial^2 K}{\partial \bar b \partial B} = 0\,, \qquad
\frac{\partial^2 K}{\partial \bar a \partial A}
+ \frac{\partial^2 K}{\partial b \partial \bar B} = 0\,.
\label{p1}
\eea
Analogously, from Eqs. (\ref{x2})\,, (\ref{x3}) one finds
further constraints on $K$
\bea
\frac{\partial^2 K}{\partial a \partial B} = 0\,, \qquad
\frac{\partial^2 K}{\partial \bar b \partial B} = 0\,, \qquad
\frac{\partial^2 K}{\partial a \partial \bar A} = 0\,, \qquad
\frac{\partial^2 K}{\partial \bar b \partial \bar A} = 0\,,\nn\\
\frac{\partial^2 K}{\partial \bar a \partial \bar B} = 0\,, \qquad
\frac{\partial^2 K}{\partial b \partial \bar B} = 0\,, \qquad
\frac{\partial^2 K}{\partial \bar a \partial A} = 0\,, \qquad
\frac{\partial^2 K}{\partial b \partial A} = 0\,
\label{p2}
\eea
and, in addition, the same two Laplace equations as in the first line of (\ref{p1})\,.

To find a solution to these constraints, it is convenient to introduce
doublets of ${\cal N}{=}(2,2)$ superfields as follows
\be
a^{\alpha} = (a\,, \bar b)\,, \quad  \ A^{\alpha} = (\bar A\,, B)\,, \quad
\bar a^{\alpha} = (\bar a\,, b)\,, \quad \bar A^{\alpha} = (A\,, \bar B)
\label{doub1}
\ee
where $\alpha$\,, $\beta = 1,2$\,.
With this new notation the set of Eqs. (\ref{p1})\,, (\ref{p2}) takes the more
concise form
\bea
\frac{\partial^2 K}{\partial A \partial \bar A}
+ \frac{\partial^2 K}{\partial B \partial \bar B} = 0\,, \qquad
&&\frac{\partial^2 K}{\partial a \partial \bar a}
+ \frac{\partial^2 K}{\partial b \partial \bar b} = 0\,,
\label{M0}\\
\frac{\partial^2 K}{\partial a^{\alpha} \partial A^{\beta}} = 0\,,\qquad
&&
\frac{\partial^2 K}{\partial \bar a^{\alpha} \partial \bar A^{\beta}} = 0\,,
\label{M1} \\
\eps^{\alpha\, \beta}
\frac{\partial^2 K}{\partial a^{\alpha} \partial \bar A^{\beta}} = 0\,, \qquad
&&
\eps^{\alpha\, \beta}
\frac{\partial^2 K}{\partial \bar a^{\alpha} \partial A^{\beta}} = 0\,.
\label{M2}
\eea
The solution of equations (\ref{M1})\,, (\ref{M2}) is as follows
(see the Appendix B for details)
\be
K(A^{\alpha}\,, \bar A^{\alpha}\,, a^{\alpha}\,, \bar a^{\alpha})
\,=\, T(A^{\alpha}\,, \bar A^{\alpha}) + h(a^{\alpha}\,, \bar a^{\alpha})\,.
\label{GEN}
\ee
In addition, each term in the r.h.s. of \p{GEN} obeys its own
four-dimensional Laplace equation, so the constraints \p{M0} 
are also satisfied.

\subsection{Action for dual twisted multiplets}

In the general case the `test' action of multiplets $\hat q^{\,\und i\, a}$
and $\hat q^{\,i\, \und a}$ can be written in the superspace 
${\bf R}^{(1,1|2,2)}$ as
\be
S_{(4,4)} = \int \mu \, K(a\,,\, \bar a\,,\, b\,,\, \bar b\,,\, 
\alpha\,,\, \bar \alpha\,,\,
\beta\,,\, \bar \beta)\,.
\label{Lagr2}
\ee
The action (\ref{Lagr2}) is invariant under both manifest ${\cal N}{=}(2,2)$ 
supersymmetry and generalized K\"ahler gauge transformations
\be
\delta K = f\,(a\,,b\,, \alpha\,, \beta)
+ g\,(a\,, \bar b\,, \alpha\,, \bar \beta)
+ \bar f\,(\bar a\,, \bar b\,, \bar \alpha\,, \bar \beta)
+ \bar g\,(\bar a\,, b\,, \bar \alpha\,, \beta)\,.
\ee
As in the previous case, the requirement that this action possesses
an extra ${\cal N}{=}(2,2)$ supersymmetry
leads to some differential constraints on the function $K\,$. To find them 
we exploit the general invariance condition (\ref{deltaS}) (denoting 
the relevant functions by the same letters) and the superfield 
transformation laws (\ref{s2})\,. The resulting constraints are
\bea
\eps^{\,+} \quad \Rightarrow \quad
\frac{\partial K}{\partial \bar b} = \frac{\partial F}{\partial a}\,, \quad
\frac{\partial K}{\partial \bar a} = - \frac{\partial F}{\partial b}\,, \quad
\frac{\partial K}{\partial \beta} = \frac{\partial G}{\partial \bar \alpha}\,, \quad
\frac{\partial K}{\partial \alpha} = - \frac{\partial G}{\partial \bar \beta}\,,
\nn\\
\bar \eps^{\,+} \quad \Rightarrow \quad
\frac{\partial K}{\partial b} = \frac{\partial \bar F}{\partial \bar a}\,, \quad
\frac{\partial K}{\partial a} = - \frac{\partial \bar F}{\partial \bar b}\,, \quad
\frac{\partial K}{\partial \bar \beta} = \frac{\partial \bar G}{\partial \alpha}\,, \quad
\frac{\partial K}{\partial \bar \alpha} = - \frac{\partial \bar G}{\partial \beta}\,,
\nn\\
\eps^{\,-} \quad \Rightarrow \quad
\frac{\partial K}{\partial \bar b} = - \frac{\partial P}{\partial \bar a}\,, \quad
\frac{\partial K}{\partial a} = \frac{\partial P}{\partial b}\,, \quad
\frac{\partial K}{\partial \beta} = - \frac{\partial H}{\partial \alpha}\,, \quad
\frac{\partial K}{\partial \bar \alpha} = \frac{\partial H}{\partial \bar \beta}\,,
\nn\\
\bar \eps^{\,-} \quad \Rightarrow \quad
\frac{\partial K}{\partial b} = - \frac{\partial \bar P}{\partial a}\,, \quad
\frac{\partial K}{\partial \bar a} = \frac{\partial \bar P}{\partial \bar b}\,, \quad
\frac{\partial K}{\partial \bar \beta} = - \frac{\partial \bar H}{\partial \bar \alpha}\,, \quad
\frac{\partial K}{\partial \alpha} = \frac{\partial \bar H}{\partial \beta}\,.
\label{y}
\eea
In addition to these equations, there arise analyticity-type conditions 
for the functions in the r.h.s. of Eq.~(\ref{deltaS})
\bea
\eps^{\,+} \quad \Rightarrow \quad
\frac{\partial G}{\partial \bar a} = 0\,, \quad
\frac{\partial G}{\partial \bar b} = 0\,, \quad
\frac{\partial F}{\partial \alpha} = 0\,, \quad
\frac{\partial F}{\partial \beta} = 0\,,
\nn\\
\bar \eps^{\,+} \quad \Rightarrow \quad
\frac{\partial \bar G}{\partial a} = 0\,, \quad
\frac{\partial \bar G}{\partial b} = 0\,, \quad
\frac{\partial \bar F}{\partial \bar \alpha} = 0\,, \quad
\frac{\partial \bar F}{\partial \bar \beta} = 0\,,
\nn\\
\eps^{\,-} \quad \Rightarrow \quad
\frac{\partial H}{\partial a} = 0\,, \quad
\frac{\partial H}{\partial \bar b} = 0\,, \quad
\frac{\partial P}{\partial \bar \alpha} = 0\,, \quad
\frac{\partial P}{\partial \beta} = 0\,,
\nn\\
\bar \eps^{\,-} \quad \Rightarrow \quad
\frac{\partial \bar H}{\partial \bar a} = 0\,, \quad
\frac{\partial \bar H}{\partial b} = 0\,, \quad
\frac{\partial \bar P}{\partial \alpha} = 0\,, \quad
\frac{\partial \bar P}{\partial \bar \beta} = 0\,.
\label{yy}
\eea
From Eqs.~(\ref{y})\,, (\ref{yy}) one finds that the potential $K$
satisfies the following integrability conditions
\bea
\frac{\partial^2 K}{\partial \bar a \partial \beta} = 0\,, \quad
\frac{\partial^2 K}{\partial \bar b \partial \beta} = 0\,, \quad
\frac{\partial^2 K}{\partial \bar a \partial \alpha} = 0\,, \quad
\frac{\partial^2 K}{\partial \bar b \partial \alpha} = 0\,,\nn\\
\frac{\partial^2 K}{\partial a \partial \bar \beta} = 0\,, \quad
\frac{\partial^2 K}{\partial b \partial \bar \beta} = 0\,, \quad
\frac{\partial^2 K}{\partial a \partial \bar \alpha} = 0\,, \quad
\frac{\partial^2 K}{\partial b \partial \bar \alpha} = 0\,,\nn\\
\frac{\partial^2 K}{\partial a \partial \beta} = 0\,, \quad
\frac{\partial^2 K}{\partial \bar b \partial \bar \alpha} = 0\,,\quad
\frac{\partial^2 K}{\partial \bar a \partial \bar \beta} = 0\,, \quad
\frac{\partial^2 K}{\partial b \partial \alpha} = 0\,.
\label{R}
\eea
Besides this, the potential $K$ should also obey two independent Laplace
equations
\be
\frac{\partial^2 K}{\partial a \partial \bar a}
+ \frac{\partial^2 K}{\partial b \partial \bar b} = 0\,, \quad
\frac{\partial^2 K}{\partial \alpha \partial \bar \alpha}
+ \frac{\partial^2 K}{\partial \beta \partial \bar \beta} = 0\,.
\label{N}
\ee
Introducing the doublet notation for superfields like in the previous case
\be
a^{\alpha} = (a\,, \bar b)\,, \quad \Gamma^{\alpha} = (\bar \alpha\,, \beta)\,, \quad
\bar a^{\alpha} = (\bar a\,, b)\,, \quad \bar \Gamma^{\alpha} = (\alpha\,, \bar \beta)\,,
\label{doub2}
\ee
we cast the set (\ref{R}) in the following compact form
\be
\frac{\partial^2 K}{\partial a^{\alpha} \partial \Gamma^{\beta}} = 0\,, \quad
\frac{\partial^2 K}{\partial \bar a^{\alpha} \partial \bar \Gamma^{\beta}} = 0\,,
\label{N1}
\ee
\be
\eps^{\alpha\, \beta}
\frac{\partial^2 K}{\partial a^{\alpha} \partial \bar \Gamma^{\beta}} = 0\,, \quad
\eps^{\alpha\, \beta}
\frac{\partial^2 K}{\partial \bar a^{\alpha} \partial \Gamma^{\beta}} = 0
\label{N2}
\ee
(no summation over repeating indices)\,. Eqs.~(\ref{N})\,, (\ref{N1})\,, (\ref{N2})
form the full set of
differential constraints on the function $K\,$.
The solution of Eqs.~(\ref{N1})\,, (\ref{N2}) shows up the same separation as
in the previous case (i.e., for a pair of non-dual ${\cal N}{=}(4,4)$ twisted multiplets)
\be
K(a^{\alpha}\,, \bar a^{\alpha}\,, \Gamma^{\alpha}\,, \bar \Gamma^{\alpha})
\,=\, T(\Gamma^{\alpha}\,, \bar \Gamma^{\alpha}) + h(a^{\alpha}\,, \bar a^{\alpha})\,.
\ee
Then Eqs.~(\ref{N}) imply that $T$ and $h$ should satisfy their own Laplace
equations.

Thus we have demonstrated that the general sigma model actions of
both `self-dual' and `non-self-dual' pairs of the twisted 
${\cal N}{=}(4,4)$ multiplets are reduced to the proper sums of 
${\cal N}{=}(4,4)$ supersymmetric actions of single multiplets. 
In other words, each action in a sum involves only those ${\cal N}{=}(2,2)$
chiral and twisted chiral superfields which belong to one 
${\cal N}{=}(4,4)$ twisted multiplet.
From these results it follows that chiral and twisted chiral 
${\cal N}{=}(2,2)$ superfields belonging to different ${\cal N}{=}(4,4)$ 
twisted multiplets and hence having different transformation properties
under the hidden ${\cal N}{=}(2,2)$ supersymmetry cannot interact 
with each other via the sigma-model type actions.

Our ${\cal N}{=}(2,2)$ superspace analysis confirms the results
obtained for different types of ${\cal N}{=}(4,4)$ twisted multiplets
in the bi--harmonic superspace approach~\cite{IS1}\,.
In that paper the separation property was proved for the actions depending on
arbitrary number of ${\cal N}{=}(4,4)$\, twisted multiplets of various kind.
The above analysis can be also extended to this general case,
with the same ultimate conclusions.

\subsection{Potential terms}

In Sect. III, on the example of the twisted multiplets $\hat q^{\, i\, a}\,$,
we showed how to construct ${\cal N}{=}(4,4)$ supersymmetric potential (or mass)
terms for such multiplets in the superspace ${\bf R}^{(1,1|2,2)}\,$.
In our previous paper~\cite{IS1} we found that for the ${\cal N}{=}(4,4)$
twisted multiplets belonging to a `self-dual' pair one can write invariant
mixed mass terms. These terms are in fact of the same form as
those given in~\cite{IK, IKL, GI, G1}\,.

The results of Ref.~\cite{IS1} can be easily reformulated in terms of
the ${\cal N}{=}(2,2)$ chiral and twisted chiral superfields.
Indeed, for a `self-dual' pair of the twisted multiplets, e.g.,
$\hat q^{\,i\, \und a}$ and $\hat q^{\,\und i\,a}\,$, the corresponding
`test' potential term bilinear in superfields can be written as
\be
S^{\,M}_{(4,4)} \,=\, i\, M \int \mu\, \Big \{\, l\,(\theta^+ \theta^-) \,
\bar \alpha\, \bar a
+ k\, (\bar \theta^+ \bar \theta^-)\, \alpha\, a  
+ n\, (\theta^+ \bar \theta^-)\, \bar \beta\,
\bar b + p\, (\bar \theta^+ \theta ^-) \,\beta\, b\, \Big \}\,,
\label{mass2}
\ee
while for the case of non-self-dual pair, e.g.,  $\hat q^{\,i\, a}$ and
$\hat q^{\,i\, \und a}\,$, it has the following form
\be
S^{\,\tilde {M}}_{(4,4)} \,=\, i\, \tilde {M} \int \mu\, 
\Big \{\, l^{\prime}\,(\theta^+ \theta^-)\, \bar a\, \bar A
+ k^{\prime}\, (\bar \theta^+ \bar \theta^-)\,  a\, A 
+ n^{\prime}\, (\theta^+ \bar \theta^-)\, \bar b\,
\bar B + p^{\prime}\, (\bar \theta^+ \theta ^-) \,b\, B \Big \}\,.
\label{mass3}
\ee
In Eqs.~(\ref{mass2})\,, (\ref{mass3})\,, $l\,,l^{\prime}$\, etc. are 
some numerical coefficients unspecified
for the moment. These coefficients have the same meaning as 
the harmonic constants $C^{\,p,q}$ with
the $U(1)\times U(1)$ charges $p$ and $q$  introduced in~\cite{IS1}\,.
Computing the variation of the action (\ref{mass2}) with respect 
to the extra ${\cal N}{=}(2,2)$ supersymmetry, it is easy to find that 
(\ref{mass2}) is invariant provided that these constants are equal
\be
l=k=n=p\,.\label{COND}
\ee
At the same time, requiring the potential action (\ref{mass3}) of 
non-dual multiplets to be invariant under this extra ${\cal N}{=}(2,2)$ 
supersymmetry implies the relevant constants to identically vanish
\be
l^{\prime}=k^{\prime}=n^{\prime}=p^{\prime}=0\,,
\ee
which forbids mixed mass terms for such a pair. Analogous results were 
obtained in~\cite{IS1}\,, where the harmonic constants in the `probe' 
potential terms for a pair of non-self-dual twisted multiplets were found 
to vanish as the result of imposing the requirement of ${\cal N}{=}(4,4)$ 
supersymmetry, whereas in the mass terms for a `self-dual' pair of 
the twisted multiplets these constants proved to be non-vanishing, 
with properly constrained dependence on harmonics.

Note that the most general ${\cal N}{=}(4,4)$ supersymmetric off-shell 
potential term is a sum of the mixed terms \p{mass2} with the condition 
\p{COND} and the linear terms \p{mass1} (for each twisted multiplet involved)\,. 
A net effect of eliminating the auxiliary fields in the full component action 
is the generation of some potential and mass terms for the physical bosonic 
fields~\cite{IS1} (plus some Yukawa-type couplings of physical fermions)\,.

\section{Conclusions}

In this paper we presented the description of four different types of 
${\cal N}{=}(4,4)$ twisted multiplets in ${\cal N}{=}(2,2)$, $2D$ 
superspace ${\bf R}^{(1,1|2,2)}\,$.
We showed that each type amounts off-shell to a pair of chiral and 
twisted chiral ${\cal N}{=}(2,2)$ superfields, with essentially different 
transformation properties under the extra ${\cal N}{=}(2,2)$ supersymmetry 
which completes the manifest one to the
entire ${\cal N}{=}(4,4)$ supersymmetry.
The general off-shell sigma model action for the ${\cal N}{=}(4,4)$ 
twisted multiplet of any fixed kind
can be written as an ${\bf R}^{(1,1|2,2)}$ integral of real functions 
$K$ which depend on the relevant pairs of the ${\cal N}{=}(2,2)$ 
superfields and are subjected to some differential
constraints. These constraints have the same form for every type
of the twisted multiplet and ensure the corresponding sigma model 
actions to exhibit ${\cal N}{=}(4,4)$ supersymmetry.
We also showed how the requirement of extra ${\cal N}{=}(2,2)$ 
supersymmetry constrains the potential (or mass) terms $P_1\,,P_2\,$.

We demonstrated that in more general cases, when the superpotential $K$ 
depends on ${\cal N}{=}(2,2)$ chiral and twisted chiral superfields 
belonging to different ${\cal N}{=}(4,4)$ twisted multiplets, 
the extra ${\cal N}{=}(2,2)$ supersymmetry requires the general sigma model 
action to split into to a sum of sigma model actions for separate multiplets.
The only possibility to arrange mutual interactions of
the twisted multiplets of different types is via the appropriate
invariant mixed mass terms.
The latter are bilinear in the chiral and twisted chiral superfields
belonging to a `self-dual' pair of the ${\cal N}{=}(4,4)$ twisted multiplets.
The multiplets from different such pairs can interact with each other neither
via sigma model actions nor via mass terms.

To summarize, the analysis performed in the present paper in the standard
${\cal N}{=}(2,2)\,$, $2D$  superfield formalism revealed a full agreement 
with the one given in Ref.~\cite{IS1} for different types of 
${\cal N}{=}(4,4)$ twisted multiplets within the bi-harmonic 
$SU(2)\times SU(2)$ superspace approach. The ${\cal N}{=}(2,2)\,$, $2D$ 
superfield description of all types of ${\cal N}{=}(4,4)$ twisted multiplets 
developed here can find applications in many physical and geometric problems 
to which these multiplets are relevant.

\section*{Acknowledgements}
We acknowledge a partial support from INTAS grant, project No 00-00254,
and RFBR grant, project  No 03-02-17440. The work of E.I. was
also supported by the RFBR-DFG grant No 02-02-04002, and a grant of the
Heisenberg-Landau program.

\section*{Appendix A. ${\cal N}{=}(2,2)$, $2D$ spinor derivatives and \\
constraints}

\setcounter{equation}{0}
\renewcommand{\theequation}{A.\arabic{equation}}

Here we give some details of our notations for spinor derivatives.

Starting from the quartet notation, one can define two different types 
of covariant derivatives
in the left and right light-cone coordinate sectors, such that they are 
doublets with respect
to different automorphism groups $SU(2)$ (these groups form, respectively, 
$SO(4)_L$ and $SO(4)_R$\,)
\bea
D_{i\, \und k} &=& (D_{i\, \und 1}\,, D_{i\, \und 2}) \equiv (D_{+i}\,, \bar D_{+i}) \nn\\
&=& (D_{1\, \und k}\,, D_{2\, \und k}) \equiv (D_{+\und k}\,, \bar D_{+\und k})\,,
\label{first}
\eea
\bea
D_{a\, \und b} &=& (D_{a\, \und 1}\,, D_{a\, \und 2}) \equiv (D_{-a}\,, \bar D_{-a}) \nn\\
&=& (D_{1\, \und b}\,, D_{2\, \und b}) \equiv (D_{-\und b}\,, \bar D_{-\und b})\,.
\label{fir}
\eea
The ${\cal N}{=}(2,2)$ spinor derivatives $D$ and $d$ which correspond to
the manifest and hidden supersymmetry, respectively, are defined by
\be
D_{+i} = (D_+\,, d_+) = D_{i\, \und 1}\,, \quad
\bar D^i_+ = (\bar D_+\,, \bar d_+) = \eps^{i\,k}\, \bar D_{+k}
= \eps^{i\,k}\, D_{k\, \und 2}\,,
\label{firstx}
\ee
\be
D_{-a} = (D_-\,, d_-) = D_{a\, \und 1}\,, \quad
\bar D^a_- = (\bar D_-\,, \bar d_-) = \eps^{a\,b}\, \bar D_{-b}
= \eps^{a\,b}\, D_{b\, \und 2}\,.
\label{firx}
\ee
The relations between the ${\cal N}{=}(2,2)$\, spinor derivatives
$D$\,, $d$\, and the $SU(2)$-doublet ones with the underlined indices
defined in the second lines of \p{first}, \p{fir}  are as follows
\be
D_{+\und k} = D_{1\, \und k} = (D_+\,, \bar d_+)\,, \quad
\bar D_+^{\und i} = \eps^{\und i\, \und k}\, \bar D_{\und k} 
= (\bar D_+\,, d_+)\,,
\ee
\be
D_{-\und b} = D_{1\, \und b} = (D_-\,, \bar d_-)\,, \quad
\bar D_-^{\und a} = \eps^{\und a\, \und b}\, \bar D_{\und b} 
= (\bar D_-\,, d_-)\,.
\ee
The explicit form of the covariant spinor derivatives as differential
operators in the left sector of ${\bf R}^{(1,1|4,4)}$ and 
${\bf R}^{(1,1|2,2)}$ is
\bea
&& D_{+i} = \frac{\partial}{\partial \theta^{+i}} 
+ i\bar \theta^+_i \partial_{++}\,, \quad
\bar D^i_+ = - \frac{\partial}{\partial \bar \theta^+_i} 
- i\theta^{+i} \partial_{++}\,, \nn\\
&& D_+ = \frac{\partial}{\partial \theta^+} 
+ i\bar \theta^+ \partial_{++}\,, \quad \;\;
\bar D_+ = - \frac{\partial}{\partial \bar \theta^+} 
- i\theta^+ \partial_{++}\,, \nn\\
&& d_+ = \frac{\partial}{\partial \eta^+} 
+ i\bar \eta^+\, \partial_{++}\,, \qquad \quad
\bar d_+ = - \frac{\partial}{\partial \bar \eta^+} 
- i\eta^+\, \partial_{++}\,.
\eea
The analogous expressions in the right sector are
\bea
\label{der}
&&
D_{-a} = \frac{\partial}{\partial \theta^{-a}} 
+ i\bar \theta^-_a \partial_{--}\,, \quad
\bar D^a_{-} = - \frac{\partial}{\partial \bar \theta^-_a} 
- i\theta^{-a} \partial_{--}\,, \nn\\
&&
D_- = \frac{\partial}{\partial \theta^-} 
+ i\bar \theta^- \partial_{--}\,, \quad \;\;\;
\bar D_- = - \frac{\partial}{\partial \bar \theta^-} 
- i\theta^- \partial_{--}\,, \nn\\
&&
d_- = \frac{\partial}{\partial \xi^-} + i\bar \xi^- \partial_{--}\,, \quad \qquad \;\;
\bar d_- = - \frac{\partial}{\partial \bar \xi^-} - i\xi^- \partial_{--}\,.
\eea

The ${\cal N}{=}(4,4)$ supersymmetry generators in ${\bf R}^{(1,1|4,4)}$ read
\bea
&& Q_{+i} = i\, \frac{\partial}{\partial \theta^{+i}} 
+ \bar \theta^+_i \,\partial_{++}\,, \qquad
\bar Q^i_{+} = -i\, \frac{\partial}{\partial \bar \theta^+_i} 
- \theta^{+i}\, \partial_{++}\,, \nn\\
&& Q_{-a} = i\, \frac{\partial}{\partial \theta^{-a}} 
+ \bar \theta^-_a\, \partial_{--}\,, \qquad
\bar Q^a_{-} = -i\, \frac{\partial}{\partial \bar \theta^-_a} 
- \theta^{-a}\, \partial_{--}\,.
\label{gen}
\eea

We denote the first components of the expansions of ${\cal N}{=}(4,4)$ 
superfields in (\ref{field2})--(\ref{field4}) with respect to 
the extra Grassmann coordinates $\eta$'s and $\xi$'s  as
\be
\left. {\bf a} \right |_{\eta=\xi=0}= a\,, \quad
\left. \bar {\bf a} \right |_{\eta=\xi=0}= \bar a \,, \quad
\left. {\bf b} \right |_{\eta=\xi=0}= b\,, \quad
\left. \bar {\bf b} \right |_{\eta=\xi=0}= \bar b\,,
\label{f2}
\ee
\be
\left. {\bf \cal A} \right |_{\eta=\xi=0} = \alpha \,, \quad
\left. \bar {\bf \cal A} \right |_{\eta=\xi=0}= \bar \alpha \,, \quad
\left. {\bf \cal B} \right |_{\eta=\xi=0}= \beta\,, \quad
\left. \bar {\bf \cal B} \right |_{\eta=\xi=0}= \bar \beta\,,
\label{f3}
\ee
\be
\left. {\sf A} \right |_{\eta=\xi=0}= {\sf a}\,, \quad
\left. \bar {\sf A} \right |_{\eta=\xi=0}= \bar {\sf a} \,, \quad
\left. {\sf B} \right |_{\eta=\xi=0}= {\sf b}\,, \quad
\left. \bar {\sf B} \right |_{\eta=\xi=0}= \bar {\sf b}\,.
\label{f4}
\ee
In the superspace ${\bf R}^{(1,1|2,2)}$\,, these ${\cal N}{=}(2,2)$ 
superfields are subjected to the following chirality and twisted 
chirality conditions
\bea
&&
\bar D_+ a = 0\,, \qquad
\bar D_- a = 0\,, \qquad
\bar D_+ b = 0\,,\qquad
D_- b = 0\,, \nn\\
&&
D_+ \bar a = 0\,,\qquad
D_- \bar a = 0\,, \qquad
D_+ \bar b = 0\,, \qquad
\bar D_- \bar b = 0\,,
\label{ss2}
\eea
\bea
&&
\bar D_+ \alpha = 0\,,\qquad
\bar D_- \alpha = 0\,, \qquad
\bar D_+ \beta = 0\,, \qquad
D_- \beta = 0\,, \nn\\
&&
D_+ \bar \alpha = 0\,,\qquad
D_- \bar \alpha = 0\,, \qquad
D_+ \bar \beta = 0\,, \qquad
\bar D_- \bar \beta = 0\,,
\label{ss3}
\eea
\bea
&&
\bar D_+ {\sf a} = 0\,, \qquad
\bar D_- {\sf a} = 0\,, \qquad
\bar D_+ {\sf b} = 0\,,\qquad
D_- {\sf b} = 0\,, \nn\\
&&
D_+ \bar {\sf a} = 0\,,\qquad
D_- \bar {\sf a} = 0\,, \qquad
D_+ \bar {\sf b} = 0\,, \qquad
\bar D_- \bar {\sf b} = 0\,.
\label{ss4}
\eea
These conditions directly follow from the defining ${\cal N}{=}(4,4)$
constraints \p{con1}--\p{con4}\,.

\section*{Appendix B. Solving constraints for $K$}

\setcounter{equation}{0}
\renewcommand{\theequation}{B.\arabic{equation}}

Here we deduce the explicit solution of the constraints on 
the superpotential $K$ which involves ${\cal N}{=}(4,4)$ twisted 
multiplets of two different types
belonging to a `non-self-dual' pair. These constraints are given by
Eqs.~(\ref{M1})\,, (\ref{M2}):
\be
\frac{\partial^2 K}{\partial a^{\alpha} \partial A^{\beta}} = 0\,,
\quad
\frac{\partial^2 K}{\partial \bar a^{\alpha} \partial \bar A^{\beta}} = 0\,.
\label{K1}
\ee
The doublet notation was explained in (\ref{doub1})\,.

As a first step, we partly solve (\ref{K1}) by introducing the complex
quantity
\be
F_{\beta}\, (\bar a^{\alpha}\,, A^{\alpha}\,, \bar A^{\alpha})
\equiv \frac{\partial K}{\partial A^{\beta}}\,, \quad
\bar F_{\beta}\, (a^{\alpha}\,, A^{\alpha}\,, \bar A^{\alpha})
\equiv \frac{\partial K}{\partial \bar A^{\beta}}\,.
\label{def1}
\ee
From the definition of $F_{\alpha}$ and $\bar F_{\alpha}$ one derives
the integrability conditions
\be
\frac{\partial F_{\alpha}}{\partial \bar A^{\beta}}
- \frac{\partial \bar F_{\beta}}{\partial A^{\alpha}} = 0\,.
\label{i}
\ee
Acting on this equation by the operator $\frac{\partial}{\partial a^{\rho}}$ 
and again using (\ref{def1})\,, one obtains
\be
\frac{\partial^2 \bar F_{\beta}}{\partial A^{\alpha} \partial a^{\rho}} = 0\,,
\ee
which implies
\be
\frac{\partial \bar F_{\alpha}}{\partial A^{\beta}}
= G_{\alpha \beta}\, (A^{\rho}, \bar A^{\rho})\,.
\label{g1}
\ee
Analogously, for $F_{\alpha}$ we find
\be
\frac{\partial F_{\alpha}}{\partial \bar A^{\beta}}
= \bar G_{\alpha \beta}\, (A^{\rho}, \bar A^{\rho})\,.
\label{g2}
\ee
Integrating Eqs.~(\ref{g1}) and (\ref{g2})\,, we find the following 
general solution for $F_{\alpha}$ and $\bar F_{\alpha}$
\bea
&&
F_{\alpha}\,(\bar a^{\beta}\,, A^{\beta}\,, \bar A^{\beta})
= f_{\alpha}\,(A^{\beta}\,, \bar A^{\beta}) 
+ \bar G_{\alpha}\, (A^{\beta}\,, \bar a^{\beta})\,, \nn\\
&&
\bar F_{\alpha}\, (a^{\beta}\,, A^{\beta}\,, \bar A^{\beta})
= \bar f_{\alpha}\,(A^{\beta}\,, \bar A^{\beta}) 
+ G_{\alpha}\, (\bar A^{\beta}\,, a^{\beta})\,.
\label{ii}
\eea
Substituting this into (\ref{i})\,, one finds
\be
\frac{\partial f_{\alpha}}{\partial \bar A^{\beta}}
- \frac{\partial \bar f_{\beta}}{\partial A^{\alpha}} = 0\,.
\ee
The solution of the last equation can be easily found
\be
f_{\alpha} = \frac{\partial}{\partial A^{\alpha}}\, 
T(A^{\beta}\,,\bar A^{\beta})\,, \qquad
\bar f_{\alpha} = \frac{\partial}{\partial \bar A^{\alpha}}\, 
T(A^{\beta}\,,\bar A^{\beta})\,.
\label{iii}
\ee
Then from Eqs.~(\ref{def1})\,, (\ref{ii})\,, (\ref{iii}) one derives
\bea
&&
\frac{\partial K}{\partial A^{\alpha}} = F_{\alpha}\,(\bar a^{\beta}\,, 
A^{\beta}\,, \bar A^{\beta})
= \bar G_{\alpha}\,(A^{\beta}, \bar a^{\beta})
+ \frac{\partial}{\partial A^{\alpha}}\, T(A^{\beta}\,, \bar A^{\beta})\,, \nn\\
&&
\frac{\partial K}{\partial \bar A^{\alpha}}
= \bar F_{\alpha}\, (a^{\beta}\,, A^{\beta}\,, \bar A^{\beta})
= G_{\alpha}\,(\bar A^{\beta}\,, a^{\beta})
+ \frac{\partial}{\partial \bar A^{\alpha}}\, T(A^{\beta}\,, 
\bar A^{\beta})\,,
\eea
or
\be
\bar G_{\alpha}\,(A^{\beta}\,, \bar a^{\beta})
= \frac{\partial}{\partial A^{\alpha}}\, \Big \{K - T \Big \}\,, \qquad
G_{\alpha}\,(\bar A^{\beta}\,, a^{\beta})
= \frac{\partial}{\partial \bar A^{\alpha}}\, \Big \{K - T \Big \}\,.
\label{iiii}
\ee
These relations imply the integrability conditions
\be
\frac{\partial \bar G_{\alpha}}{\partial A^{\beta}}
- \frac{\partial \bar G_{\beta}}{\partial A^{\alpha}} = 0\,, \qquad
\frac{\partial G_{\alpha}}{\partial \bar A^{\beta}}
- \frac{\partial G_{\beta}}{\partial \bar A^{\alpha}} = 0\,,
\ee
which, in turn, give that
\be
\bar G_{\alpha} = \frac{\partial}{\partial A^{\alpha}}\, 
G(A^{\beta}\,, \bar a^{\beta})\,, \qquad
G_{\alpha} = \frac{\partial}{\partial \bar A^{\alpha}}\, 
\bar G(\bar A^{\beta}\,, a^{\beta})\,.
\label{z1}
\ee
Substituting (\ref{z1}) into (\ref{iiii}) leads to the set of equations
\be
\frac{\partial}{\partial A^{\alpha}}\,
\Big \{K -T - G(A^{\beta}\,, \bar a^{\beta}) \Big \} \,=\, 0\,, \qquad
\frac{\partial}{\partial \bar A^{\alpha}}\,
\Big \{K -T - \bar G(\bar A^{\beta}\,, a^{\beta}) \Big \} \,=\, 0\,,
\label{z2}
\ee
which can be easily solved as
\bea
&&
(i) \quad K - T \,=\, G(A^{\alpha}\,, \bar a^{\alpha})
+ \Omega(\bar A^{\alpha}\,, a^{\alpha}\,, \bar a^{\alpha})\,, \nn\\
&&
(ii) \quad K - T \,=\, \bar G(\bar A^{\alpha}\,, a^{\alpha})
+ \bar \Omega(A^{\alpha}\,, a^{\alpha}\,, \bar a^{\alpha})\,.
\label{z3}
\eea
Expressing $K-T$ from Eq.~{\it (ii)} in (\ref{z3}) and substituting it into
the first equation in (\ref{z2})\,, one finds
\be
\bar \Omega \,=\,G(A^{\alpha}\,, \bar a^{\alpha}) 
+ h(a^{\alpha}\,, \bar a^{\alpha})\,.
\ee
Analogously, expressing $K-T$ from Eq.~{\it (i)} and substituting it into
the second equation in (\ref{z2})\,, one obtains
\be
\Omega \,=\, \bar G(\bar A^{\alpha}\,, a^{\alpha}) 
+ h(a^{\alpha}\,, \bar a^{\alpha})\,.
\ee

Finally, the full solution of the constraints on 
the superpotential $K$ is a sum of four pieces
\be
\label{SOL}
K(A^{\alpha}\,, \bar A^{\alpha}\,, a^{\alpha}\,, \bar a^{\alpha})
\,=\, T(A^{\alpha}\,, \bar A^{\alpha}) + h(a^{\alpha}\,, \bar a^{\alpha})
+ G(A^{\alpha}\,, \bar a^{\alpha}) + \bar G (\bar A^{\alpha}\,, a^{\alpha})\,.
\ee
Taking into account the definition of the doublets $A^{\alpha}$, $a^{\alpha}$
and their complex conjugates, as well as the Laplace equations \p{M0}\,,
we conclude that the first two terms in (\ref{SOL})
correspond to the potentials of ${\cal N}{=}(4,4)$\, supersymmetric 
sigma model actions for two independent twisted multiplets 
$\hat q^{\,i\,a}$ and $\hat q^{\,i\, \und a}\,$.
The last two terms can be removed by the generalized K\"ahler 
gauge transformations in \p{Gauge1} corresponding to the gauge function $g$\,. 
So they do not make contribution into the ${\bf R}^{(1,1|2,2)}$ superfield action.

Thus, the final result for the potential $K$ is
\be
K(A^{\alpha}\,, \bar A^{\alpha}\,, a^{\alpha}\,, \bar a^{\alpha})
\,=\, T(A^{\alpha}\,, \bar A^{\alpha}) + h(a^{\alpha}\,, \bar a^{\alpha})\,.
\ee

The proof for the case of the chiral and twisted-chiral ${\cal N}{=}(2,2)$ 
superfields which form a `self-dual' pair of ${\cal N}{=}(4,4)$ twisted 
multiplets follows the same route. It can be also straightforwardly extended 
to the case with multiple twisted multiplets of various types.

\end{document}